\documentclass[useAMS,usenatbib]{mn2e} 
\usepackage{times}
\usepackage{graphicx}

\bibpunct{(}{)}{;}{a}{}{,}

\def\uprime{^{\prime}}
\def\Nprimetot{N{\uprime}_{\mathrm{tot}}}
\def\Ntot{N_{\mathrm{tot}}}
\def\Mtot{M_{\mathrm{tot}}}
\def\Neff{\cal{N}}

 \title[Biases in stellar population synthesis]{On biases in the predictions 
of stellar population synthesis models}
 
\author[M. Cervi\~no and D. Valls--Gabaud]
       {M. Cervi\~no$^{1,2}$ and D. Valls--Gabaud$^{3}$\\ 
     $^1$ Instituto de Astrof\'\i sica de Andalucia
          (CSIC) Camino Bajo de Hu\'etor 24, Granada 18080, Spain\\ 
     $^2$ Laboratorio de Astrof\'\i sica Espacial y F\'\i sica fundamental 
          (INTA) Apdo. 50272, Madrid 28080, Spain\\ 
     $^3$ UMR CNRS 5572, Laboratoire d'Astrophysique, Observatoire 
           Midi-Pyr\'en\'ees, 14 Avenue Edouard Belin, 31400 Toulouse,
           France}
 
\begin{document}

\date{Accepted September 16, 2002, Received August 7, 2002 ; in original
form .... }
 
\pagerange{\pageref{firstpage}--\pageref{lastpage}} \pubyear{2002}

\maketitle
 
\label{firstpage}

\begin{abstract}
Sampling fluctuations in stellar populations give rise to dispersions in
  observables when a small number of sources contribute effectively to the
  observables. This is the case for a variety of linear functions of the
  spectral energy distribution (SED) in small stellar systems, such as
  galactic and extragalactic H{\sc ii} regions, dwarf galaxies or stellar
  clusters. In this paper we show that sampling fluctuations also introduce
  systematic biases and multi-modality in non-linear functions of the SED,
  such as luminosity ratios, magnitudes and colours.  In some cases, the
  distribution functions of rational and logarithmic quantities are
  bimodal, hence complicating considerably the interpretation of these
  quantities in terms of age or evolutionary stages.  These biases can be
  only assessed by Monte Carlo simulations. We find that biases are usually
  negligible when the effective number of stars, $\Neff$, which contribute
  to a given observable is larger than 10.  Bimodal distributions may
  appear when $\Neff$ is between 10 and 0.1. Predictions from any model of
  stellar population synthesis become extremely unreliable for small
  $\Neff$ values, providing an operational limit to the applicability of
  such models for the interpretation of integrated properties of stellar
  systems.  In terms of stellar masses, assuming a Salpeter Initial Mass
  Function in the range 0.08 -- 120 M$_\odot$, $\Neff$=10 corresponds to
  about 10$^5$ M$_\odot$ (although the exact value depends on the age and
  the observable).  This bias may account, at least in part, for claimed
  variations in the properties of the stellar initial mass function in
  small systems, and arises from the discrete nature of small stellar
  populations.
\end{abstract}


\maketitle

\section{Introduction and motivation} 
 
The comparison of observations to theoretical models is one of the basic
steps which allows an understanding of natural phenomena. The comparison is
not always satisfactory and leads to either new insights on the nature of
the phenomena observed or to revisions of the theoretical framework within
which the models are made, or both.
 
In general models have an {\it intrinsic} uncertainty, in addition to
possible systematic effects, and one seeks to minimise both in order to
infer a robust interpretation of the observations. In the case of stellar
population synthesis, the former uncertainties have often been neglected in
comparison with the latter, where external errors (i.e. uncertainties on
the input assumptions) have been much discussed, e.g. \citet{Lei96} or
\citet{Bru01}.  However, the {\it predicted} integrated properties of small
stellar systems suffer from large intrinsic dispersions arising from the
very nature of the systems: sampling fluctuations from, say, the initial
mass function (IMF) will give rise to fluctuations in the properties and
hence to an intrinsic dispersion in the predictions
\citep[e.g.][]{Chi88,SF97,LM99,CLC00,Cer01,Bru02Tuc,Cer02}. This effect has
been overlooked by most synthesis codes, which usually take an analytical
IMF and produce predictions which are only valid in the limit where the
number of stars is formally infinite.

\begin{figure}
\centerline{\includegraphics[width=6cm,angle=270]{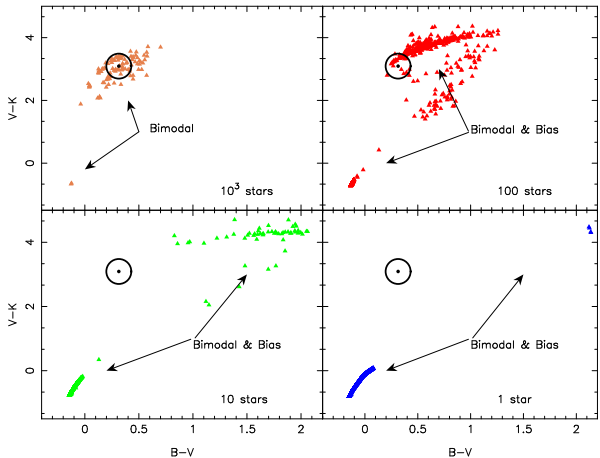}}
\caption{Integrated $B-V$ colour vs. $V-K$ colour of 10 Myr old stellar
clusters with 1000, 100, 10 and 1 star, respectively, with masses between 2
and 120 M$_\odot$ distributed following a Salpeter IMF (triangles). Each
triangle represents a Monte Carlo simulation with these parameters. The
location of a stellar population with an infinite number of stars is
indicated with the $\odot$ symbol. The colours of the sparsely-populated
clusters show a multimodal distribution in this diagram, and their average
values are biased with respect to the colours of an infinite stellar
population.}
\label{fig:GTC}
\end{figure}

In addition to this dispersion --which can be quantified and properly taken
into account with the tools developed by \citet{Cer02}-- there is also a
more subtle effect arising from this incomplete sampling.  Integrated
quantities which scale linearly with the number of stars (such as the
luminosity in a given photometric band, or the overall spectral energy
distribution, SED) can be predicted for any stellar system from the
scaled-down outputs of synthesis codes with effectively predict properties
for very large stellar systems, that is, systems which fully sample the
distribution function of masses. However, for quantities which do {\it not}
scale {\it linearly} with the number of stars, such as luminosity ratios,
magnitudes, colours, equivalent widths, etc, such a scaling cannot be
performed properly, and the only solution is to make extensive Monte Carlo
simulations. We can already foresee that biases will arise from the
incomplete sampling of the underlying distribution function. For instance,
infrared colours may be dominated, at some ages, by a handful of high
luminosity, IR-bright stars.  Monte Carlo simulations calculated by
\citet{SF97} ~show that the average colour is a function of the number of
stars in the simulation.  Similarly, the average optical colours of stellar
clusters, at a given age, are functions of the total mass in the cluster,
as shown for instance by \citet{GB93} and more recently by \citet{G00} and
\citet{Bru02Tuc}.

This is best illustrated with an example in a simple colour-colour
diagram. Figure \ref{fig:GTC} shows the point (indicated by the $\odot$
symbol) where a 10 Myr-old stellar population with an infinite number of
stars would be. Each triangle in each panel represents one Monte Carlo
simulation of a 10 Myr-old stellar population with a total number of stars
(with masses between 2 and 120 M$_\odot$ distributed with a Salpeter IMF)
of 1000, 100, 10 and 1 star, respectively\footnote{For reference,
$\Ntot$=1000 stars within 2 and 120 M$_\odot$ have a weight of $\Mtot
\approx 6 \times 10^3$ M$_\odot$, and would the IMF be extended down to
0.08 M$_\odot$, the mass of the cluster would be $M_{\rm tot}$=$2.25 \times
10^4$ M$_\odot$.}. Figure \ref{fig:GTC} shows three important effects:

{\it (a)} The distribution of colours in this diagram is clearly bimodal or
 even multimodal. The points do not cluster around a well-defined center,
 but are very scattered over the entire diagram, at odds with the
 (standard) predicted value for an infinitely-large stellar population.

{\it (b)} If the average colour is computed for clusters with the same
 number of stars, its value is very different from the one given by the
 standard prediction : the average colour is biased, and the bias depends
 on the total number of stars present in the population.

{\it (c)} The area covered by the points (each representing one Monte Carlo
 simulation) is not a monotonic function of the number of stars in the
 simulation.

Note that these effects are not limited of course to clusters of this age,
but do also appear in more evolved populations, since these effects are a
generic feature of under-sampled clusters\footnote{See for example the
analysis by \cite{Bru02Tuc} where the positions of LMC clusters are
interpreted in terms of Monte Carlo simulations of clusters of $10^5$,
$10^4$ and $10^3$ M$_\odot$ at different ages.}.

In this paper we provide a first order analytical estimation on {\it when
and why synthesis models which use an analytical formulation of the IMF are
not be able to reproduce the observed properties of stellar systems} and
therefore when Monte Carlo simulations are absolutely required in order to
make predictions.  We show that bias and multimodality features are a {\it
generic} property of non-linear functions of the SED, and will always be
present when trying to apply population synthesis models to small size
stellar populations. Their importance cannot be stressed enough: many of
the seemingly 'peculiar' properties of some systems can, at least in part,
be accounted for by this bias and multimodality effects.

This is an important issue since odd values of these properties often lead
to non standard interpretations. For instance, when colours or equivalent
widths (EWs) of small stellar systems appear to be peculiar (that is,
peculiar with respect to the values predicted by synthesis models of a
large number of stars), they are often interpreted in terms of IMF
variability : either the mass cutoffs have to be truncated, and/or the
slope of the IMF has to be changed in order to produce the 'correct'
colours or EWs.  Since the lower mass cutoff determines the overall
mass-to-light ratio, it is usually the upper mass limit which is varied to
account for the data. Examples of these interpretations abound in the
recent literature, starting perhaps with \citet{ST76} who, on the basis of
the variation of the equivalent width of H$\beta$ in H{\sc ii} regions
across M 101, inferred a dependence of the upper mass cutoff with
metallicity.  Line ratios included in classical diagnostic diagrams have
also been used (e.g.  \citealt{O97}, and \citealt{BKG99} for the softness
parameter) and seem to show that the IMF changes in extragalactic H{\sc ii}
regions \cite[but see however][ for an alternative explanation based on
sampling effects]{BK02}.  Colours, and colour-colour diagrams, sometimes
including the H$\alpha$ line, of clusters and starbursts also seem to point
to variations in the parameters of the IMF, from the LMC \citep{DH98}, to
clusters in the starburst galaxy IC10 \citep{H01}.

The common features of all these interpretations which point to a variation
of the IMF are (i) the systematic use of non-linear quantities (EWs, line
ratios, colours) and (ii) observations of small stellar systems (H{\sc ii}
regions, stellar clusters, starbursts, etc).

Given the scaling arguments explained above, it is interesting to see
whether sampling effects could account for this type of observations.  In
this paper we show that whilst the {\it average} values of the properties
(and their {\it dispersions}) that scale with the number of stars are
reasonably well reproduced by the models, non-linear functions such as
luminosity ratios or colours may be heavily biased. It is beyond the scope
of this paper to analyse in detail each and every observation, but given
our results, it appears very likely that the 'peculiar' properties observed
in some systems can be simply understood in terms of the bias and
multimodality introduced by sampling fluctuations of the IMF in these
stellar systems.  This has profound implications for the interpretation of
colours, average magnitudes and line ratios in such systems, as well as for
the universality (or otherwise) of the IMF.  These effects seem to be large
enough that most observations may point to the universality of the IMF,
despite a variety of physical conditions (e.g. \citealt{K02} for a general
review and \citealt{MHK99} for the case of starburst galaxies), and
therefore provide a unique insight on star formation processes in a variety
of environments.
 
The structure of the paper is the following: in Sec. \S2 we recall the
scaling properties of synthesis models.  In Sec. \S3 we analyse the
statical properties of functions of the SED, showing the biases and
multimodalities that appear in their distribution functions.  We discuss
these results and their implications in Sec. \S4 and \S5.

\section{On sampling and scaling in population synthesis} 
 
Population synthesis models often scale predictions to the total mass in
stars. There are at least two reasons for this. First, a (suitable) fixed
mass-to-light ratio can be achieved. Second, the predictions, which are
made for a unit star formation rate (that is, the rate at which a given gas
mass is transformed into stars), can be scaled to other rates, or,
equivalently, to other masses in gas or stars.

The underlying mathematical justification for this scaling is simple to
see. Take for example the average total luminosity in a given band $\mu_L =
<L>$. This is given by the weighted sum of the luminosities $L_i$ of the
$N_\mathrm{tot}$ stars present in a population at some given age :

\begin{equation} 
\mu_L(N_{\mathrm{tot}}) \; = \; \sum_{i=1}^{I} \, n_i L_i,  
\end{equation} 

\noindent where $I$ is the number of initial (zero-age) masses considered
such that $N_\mathrm{tot} = \sum_{i=1}^{I} \, n_i $ and $n_i$ is the
(integer) number of stars of type or class $i$.  For a sample of different
size $\Nprimetot$, all other parameters kept the same (age, metallicity,
etc), the following scaling applies:
 
\begin{equation} 
\mu_L({\Ntot}) \; / \; \Ntot \; \; = \;  \; \mu_L({\Nprimetot}) \; / \;
\Nprimetot.  
\end{equation} 
 
This scaling arises because, at a fixed age, $L_i$ are constants whilst
$n_i$ follow a Poisson distribution \cite[see][]{Cer02}, and the average
value of a (weighted) sum of Poisson variables is the (weighted) sum of
average values. It is this linear scaling that allows predictions made for
a given $\Ntot$ to be scaled up or down for a different number of stars.

However, the scaling is always made with the total {\it mass} in stars, not
the actual {\it number} of stars $\Ntot$ present in a population.
Obviously, for a fixed IMF, there is a direct relation between the total
mass in stars and the number of stars, but there is an important difference
in that the total mass, $\Mtot$ is a {\it random variable} when $\Ntot$ is
fixed (and vice versa).

The PDF of $\Mtot$, as of any function of the IMF, can be obtained
analytically using the characteristic function method \cite[e.g.][]{KS77},
although the inversion can be cumbersome or indeed unworkable. For the
purposes of the present paper it is enough to show the PDF of $\Mtot$
obtained from Monte Carlo simulations.

\begin{figure*} 
\centering \includegraphics[angle=270,width=14cm]{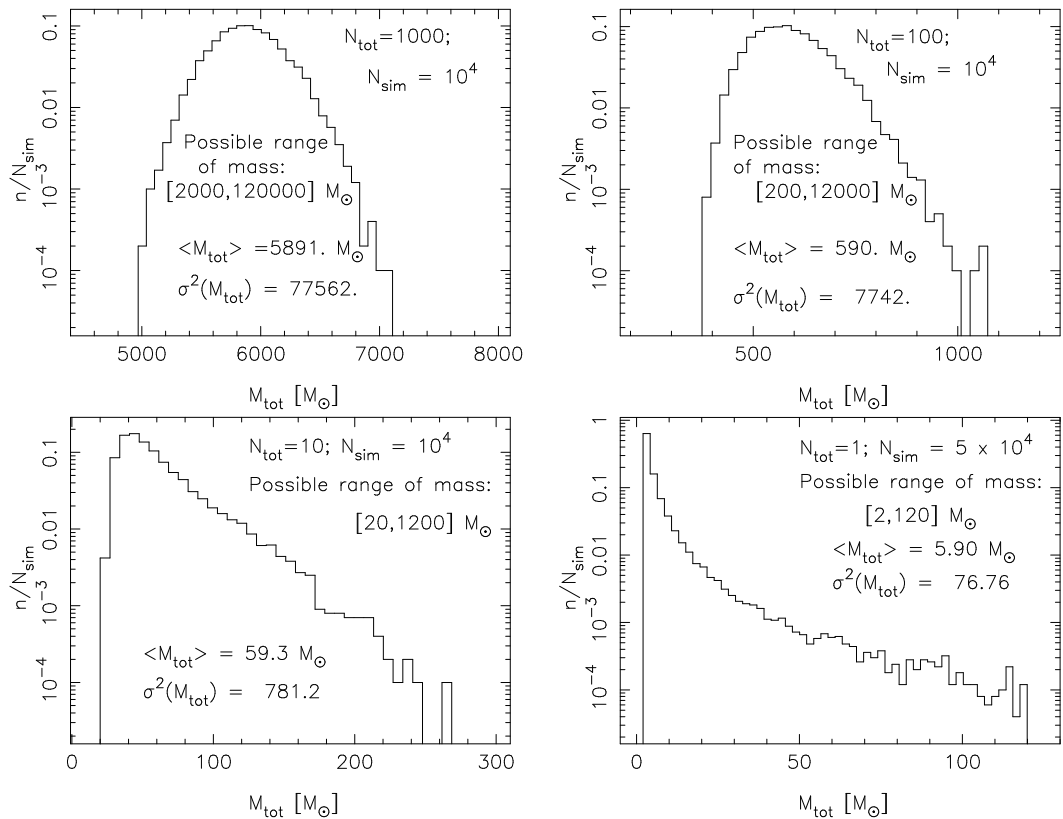}
 \caption{Probability distribution function (PDF) of the total mass $\Mtot$
 of clusters containing 1000, 100, 10 and 1 stars based on 10$^4$
 simulations ($5 \times 10^4$ for the case of 'clusters' with 1 star). The
 PDF in the case of only 1 star is trivially the IMF.}
\label{fig:IMF1} 
\end{figure*} 

For $\Ntot=1$ the PDF of $\Mtot$ is trivially the IMF itself (see
Fig. \ref{fig:IMF1} bottom right), while for large values of $\Ntot$ the
central limit theorem ensures that it converges to a Gaussian function
(Fig. \ref{fig:IMF1} top left). For intermediate values of $\Ntot$, the
distribution has a heavy tail at large values of $\Mtot$, as
Fig. \ref{fig:IMF1} shows. Note also that this tail is not sampled well
enough even though 10$^4$ simulations were made, with clusters populated by
1000 stars of masses between 2 and 120 M$_\odot$ distributed with a
Salpeter IMF. The heavy tails are a direct result of the poor sampling of
the upper mass range, where the presence of a few massive stars can be very
significant for the total mass in a small system.

The large dispersion in $\Mtot$, at $\Ntot$ fixed, shows that synthesis
models which use an analytical 'sampling' of the IMF overlook a fraction of
the intrinsic dispersion inherent to the nature of the problem: small
stellar systems will {\it always} show a dispersion in $\Mtot$ event though
$\Ntot$ is kept fixed, and inversely. So not only using $\Ntot$ is the
natural choice for the scaling, but it also leads to take into account one
of the factors which produce an intrinsic dispersion in the observables.

The evaluation of the dispersion in $\Mtot$ requires to know the
corresponding PDF, but we can estimate how it scales with $\Ntot$ from the
results of the Monte Carlo simulations. Let us assume that the IMF,
$\Phi(m)$, is given by a power law:

\begin{equation}
\frac{d N}{d m} = \Phi(m) = A m^{-\alpha},
\end{equation}

\noindent where $A$ is a normalisation constant. If the IMF is defined
between some mass cutoffs $m_\mathrm{low}$ and $m_\mathrm{up}$, the
normalisation constant is given by:

\begin{eqnarray}
A&=& \frac{(1-\alpha)}{ m_{\mathrm{up}}^{1-\alpha} -
m_{\mathrm{low}}^{1-\alpha}}.
\end{eqnarray}

In the case of only one star, the {\it average} value of the 
stellar mass is given by

\begin{eqnarray} 
<m> & = & \int_{m_\mathrm{low}}^{m_\mathrm{up}} m \,\Phi(m) \,dm = \;
 \nonumber \\ 
    & = & \frac{A}{(2-\alpha)} \, \left(m_{\mathrm{up}}^{2-\alpha} - 
 m_{\mathrm{low}}^{2-\alpha} \right) \nonumber \\ 
    & = & \frac{(1-\alpha) \, (m_{\mathrm{up}}^{2-\alpha} -
 m_{\mathrm{low}}^{2-\alpha})} {(2-\alpha) \, (m_{\mathrm{up}}^{1-\alpha} -
 m_{\mathrm{low}}^{1-\alpha})},
\end{eqnarray} 

\noindent and its variance by 

\begin{eqnarray} 
\sigma^2(m) & = & 
\int_{m_\mathrm{low}}^{m_\mathrm{up}} (m - <m>)^2 \,\Phi(m) \,dm
\nonumber \\
   &=& \frac{(1-\alpha) (m_{\mathrm{up}}^{3-\alpha} - 
m_{\mathrm{low}}^{3-\alpha})} {(3-\alpha) (m_{\mathrm{up}}^{1-\alpha} - 
m_{\mathrm{low}}^{1-\alpha})}-<m>^2.
\label{eq:MNvar} 
\end{eqnarray} 

\noindent so, square of its relative error is

\begin{eqnarray} 
\frac{\sigma(m)^2}{<m>^2} & = & 
\left[  \frac{(2-\alpha)^2 \, (m_{\mathrm{up}}^{3-\alpha} -
m_{\mathrm{low}}^{3-\alpha})}{(3-\alpha)(m_{\mathrm{up}}^{2-\alpha} -
m_{\mathrm{low}}^{2-\alpha})^2}\right] \times \nonumber \\
& &  \left[ \frac{(m_{\mathrm{up}}^{1-\alpha} -
m_{\mathrm{low}}^{1-\alpha})}{(1-\alpha)} \right]- 1.
\label{eq:MNvarrel} 
\end{eqnarray} 

For the particular case of a Salpeter IMF, with cutoffs at 2 and 120
$M_\odot$, these values are $<m>=5.897$, $\sigma(m)^2=76.28$ and
$\sigma(m)/m$ = 1.481.

In general, $\Mtot$ is proportional to $\Ntot$. It can be seen that the
variance, $\sigma(\Mtot)^2$ is also proportional to $\Ntot$.  Therefore the
relative error in $\Mtot$ is inversely proportional to the square root of
the number of stars.

For stellar populations with a large number of stars, the dispersion
becomes negligible (e.g., 0.01 per cent for a system of 2$\times 10^8$
stars) and quantities which depend linearly on the number of stars can be
normalised safely to $\Mtot$ rather than to $\Ntot$ (which becomes
difficult to measure in such cases!).

\section{Distribution functions and moments 
of observables in synthesis models} 

\begin{figure} 
\centering \includegraphics[angle=270,width=7cm]{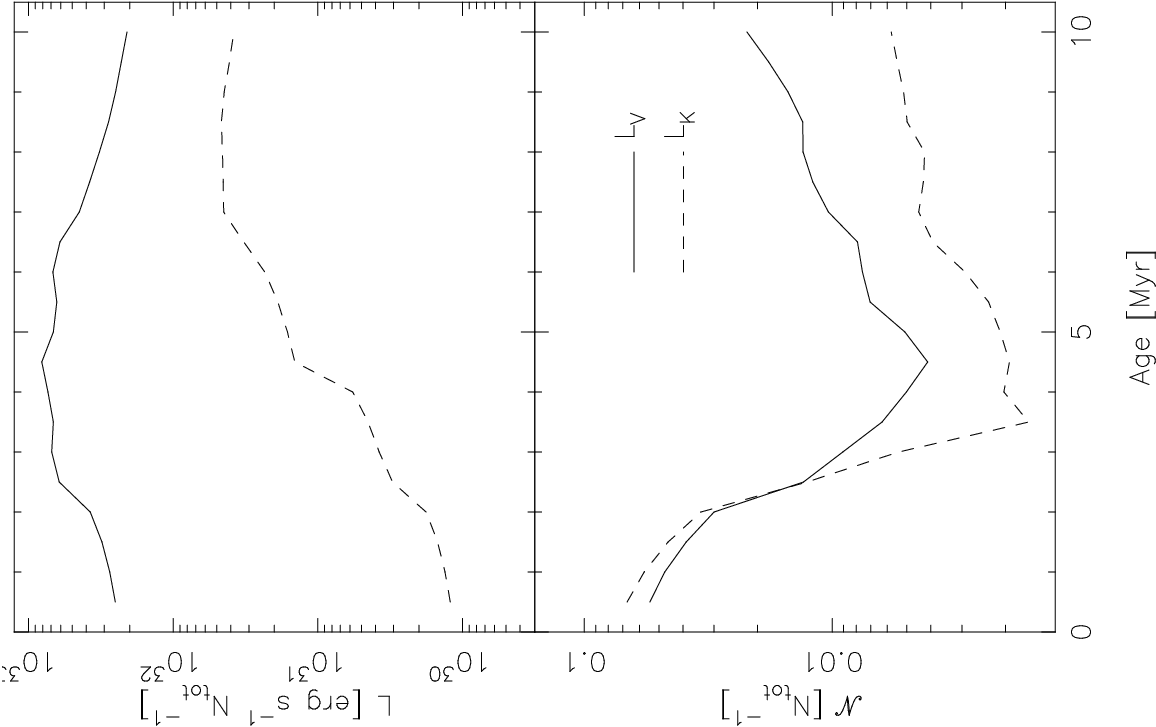}
 \caption{Temporal evolution of the average $L_V$ and $L_K$ luminosities
 and their corresponding effective number of stars, ${\cal N}$, normalised
 to the actual number of stars $\Ntot$. This is the evolution predicted for
 an infinite population of stars, fully sampling a Salpeter IMF between
 2--120 M$_\odot$.}
\label{fig:CVGLMH} 
\end{figure} 

In order to establish how relevant the effect of sampling is, we have
followed the evolution of each Monte Carlo simulation of stellar clusters
of fixed $\Ntot$, and derived their integrated luminosities in the $V$ and
$K$ bands.

Using the ensemble of simulations at fixed $\Ntot$, we have obtained mean
values and dispersions for these luminosities, their $\log$, the $L_V/L_K$
ratio and the $V-K$ colour.  To avoid introducing further parameters in
this study, we used the \cite{Kur} atmosphere models and solar metallicity
tracks with standard mass-loss rates form \cite{Schetal92}.  For the sake
of simplicity, the nebular contribution to the luminosity has not been
taken into account in these runs.

Although Monte Carlo simulations seem the natural, and only, way to
estimate the PDF of any observable, \cite{Cer02} showed that it is possible
to apply a Poissonian formalism which allows the estimation of the first
two moments of the PDF analytically. This formalism is based on the idea
that there is only an {\it effective number of stars}, ${\cal N}$, which
contribute to a given observable. Its definition is related to the
quasi-Poissonian nature of the luminosity PDF. For example, for the
observable monochromatic luminosity $L$, its effective number ${\cal N}(L)$
is given by the expression

\begin{equation} 
\frac{1}{\sqrt{{\cal N}(\mathrm{L})}} =
\frac{\sigma(\mathrm{L})}{\mu_\mathrm{L}},  
\label{eq:neff} 
\end{equation} 

\noindent as first derived by \cite{Buzz89}. 

Note that this effective number is not an actual number of stars, but
rather a measure of the weighted number of sources which contribute to a
given observable. The larger the effective number, the better determined
will the observable be.  Small values of ${\cal N}(L)$ indicate an
intrinsically large dispersion, and typically arise when a few (real) stars
dominate the observable. For instance in young clusters a few very massive
stars may dominate the overall energy budget or the SED.  In old stellar
populations, a few Asymptotic Giant Branch stars (AGB) may change
significantly the luminosity. Hence in both situations one expects an
intrinsically large dispersion in the observable, a direct result of the
discrete nature of these small populations which sample incompletely the
IMF but dominate the observable.

Figure \ref{fig:CVGLMH} shows, for reference, the time evolution of the
$L_V$ and $L_K$ luminosities (normalised to the number of
stars in the cluster), for a stellar population which fully samples the
given IMF, that is, with an infinite number of stars in the indicated mass
range, along with the evolution of their effective number of stars $\Neff$,
which readily allows an estimate of the intrinsic dispersion in these
luminosities via Eq. (\ref{eq:neff}), for any stellar population of any size
and any age. These were some of the results obtained by
\cite{Cer02}.

It is interesting to see some values of $\cal N$. From
Fig. \ref{fig:CVGLMH} can be seen that $5(2.5) \times 10^3$ stars are
needed to obtain ${\cal N}(L_K)$ (${\cal N}(L_V)$) values larger than
10. This means cluster masses around $3 (1.5) \times 10^4$ M$_\odot$ in the
mass range 2--120 M$_\odot$, and would the IMF be extended down to 0.08
M$_\odot$, the mass of the cluster would be $M_{\rm tot}$=$1.1 (0.6) \times
10^5$ M$_\odot$.

For young stellar populations, ${\cal N}$ values for different luminosities
normalised to the stellar cluster mass assuming a Salpeter IMF in the mass
range 2-120 M$_\odot$ can be found in {\tt
http://www.laeff.esa.es/users/mcs/SED/}.

In the case of old stellar populations ${\cal N}$ has been computed
directly by \cite{Buzz89}. Additional estimations can be obtained from the
evaluation of brightness fluctuations, $\bar{L}$ as shown by
\cite{Buzz93}\footnote{In the case of the work from \cite{Wor94} the
following relation applies: $M-\bar{M} = - 2.5 \log {\cal N} - 14.53$,
where $M$ is the magnitude in a given band and $\bar{M}$ is the brightness
fluctuations in magnitudes in the same band, and the term $14.53=2.5 \log
0.65\times10^6$ is used to obtain the corresponding $\cal{N}$ value
normalised to the cluster stellar mass for stars masses in the mass range
0.08--120 M$_\odot$ following a Salpeter IMF slope, instead the
normalisation used by \cite{Wor94}: 10$^6$ M$_\odot$ in the mass range
0.1--2 M$_\odot$.}.

\subsection{Linear functions} 

\begin{figure*} 
  \centering \includegraphics[angle=270,width=14cm]{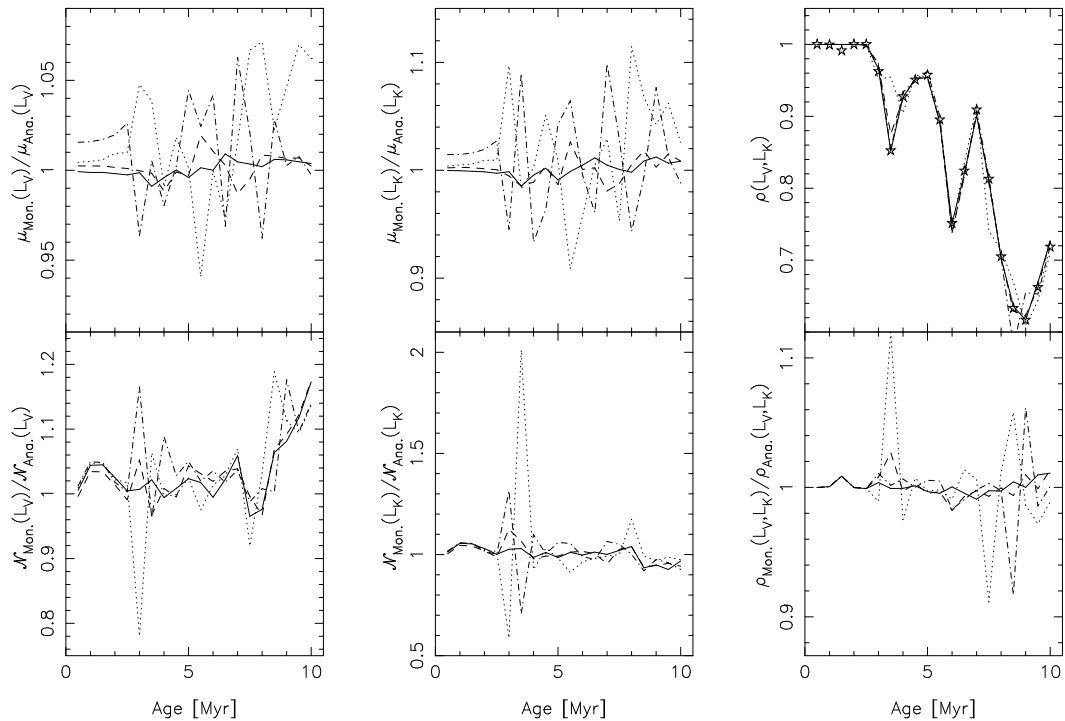}
 \caption{Comparison between Monte Carlo simulations (Mon) and
 fully-sampled (Ana) of the average values and dispersions of $L_V$ (left)
 and $L_V$ (middle).  The lines correspond to different sizes of stellar
 populations ($\Ntot$=1000 stars: solid, $\Ntot$=100 stars: dashed,
 $\Ntot$=10 stars: dot-dashed and $\Ntot$=1 star: dotted). The top panel on
 the right side give the correlation coefficient between both quantities
 (upper panel) by the simulations (lines) and the estimation proposed by
 \citet{Cer02} (stars). The ratio between the Monte Carlo and fully-sampled
 estimates are shown ion the lower panels.}
\label{fig:lum} 
\end{figure*} 

\begin{figure*} 
\centering \includegraphics[angle=270,width=12cm]{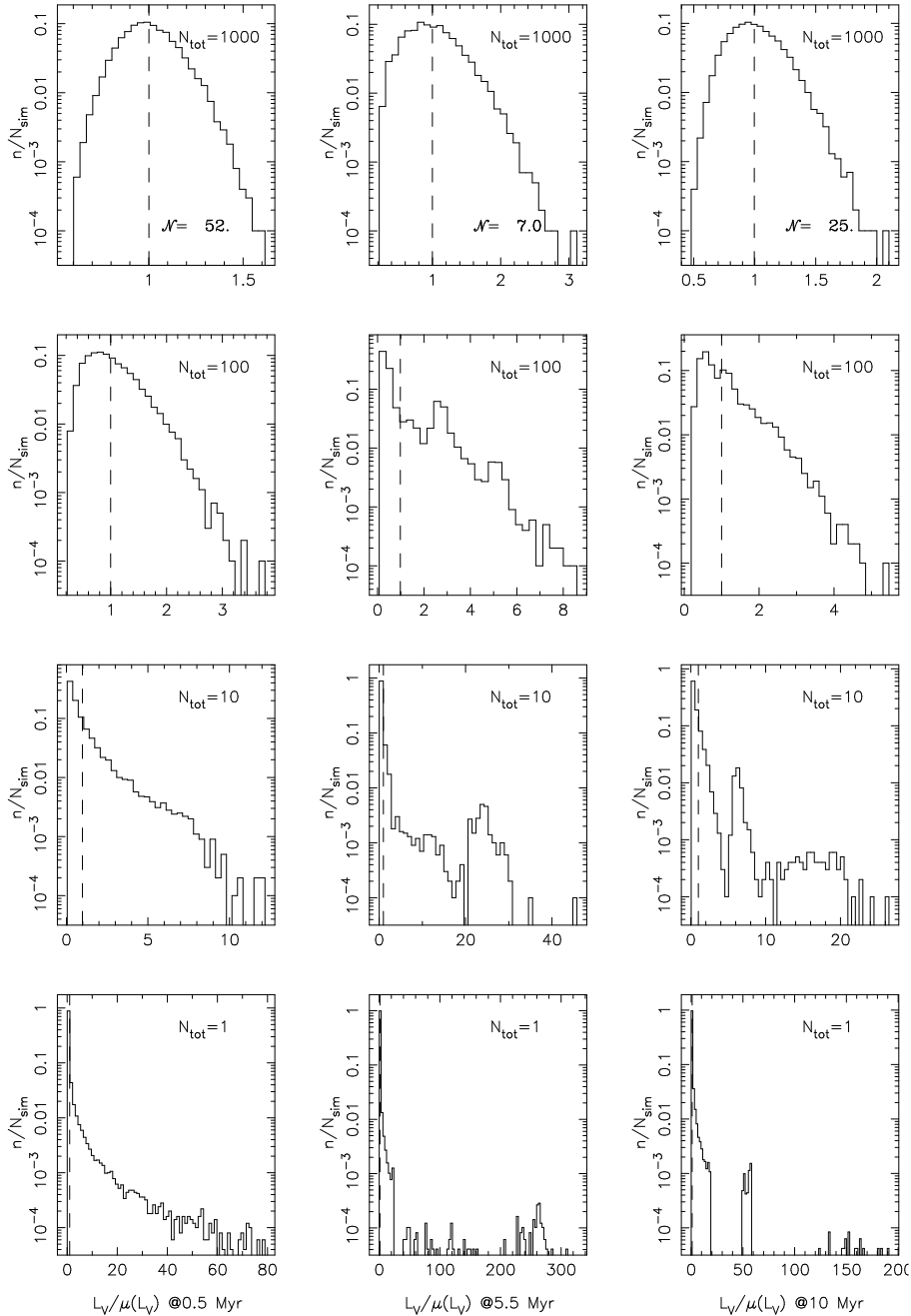} 
 \caption{PDF distributions ${\cal G}(x)$ for $x=L_V/\mu(L_V)$ at selected
 ages for a variety of sizes of stellar populations. Note that $\cal{N}$ is
 given only for clusters with $\Ntot$=1000 stars, since it scales linearly
 with $\Ntot$, that is, ${\Neff}(\Ntot) = {\Neff}(10^3) \times (\Ntot/10^3)$
 for the other populations. }
\label{fig:PDFlv} 
\end{figure*} 
 
\begin{figure*} 
\centering \includegraphics[angle=270,width=12cm]{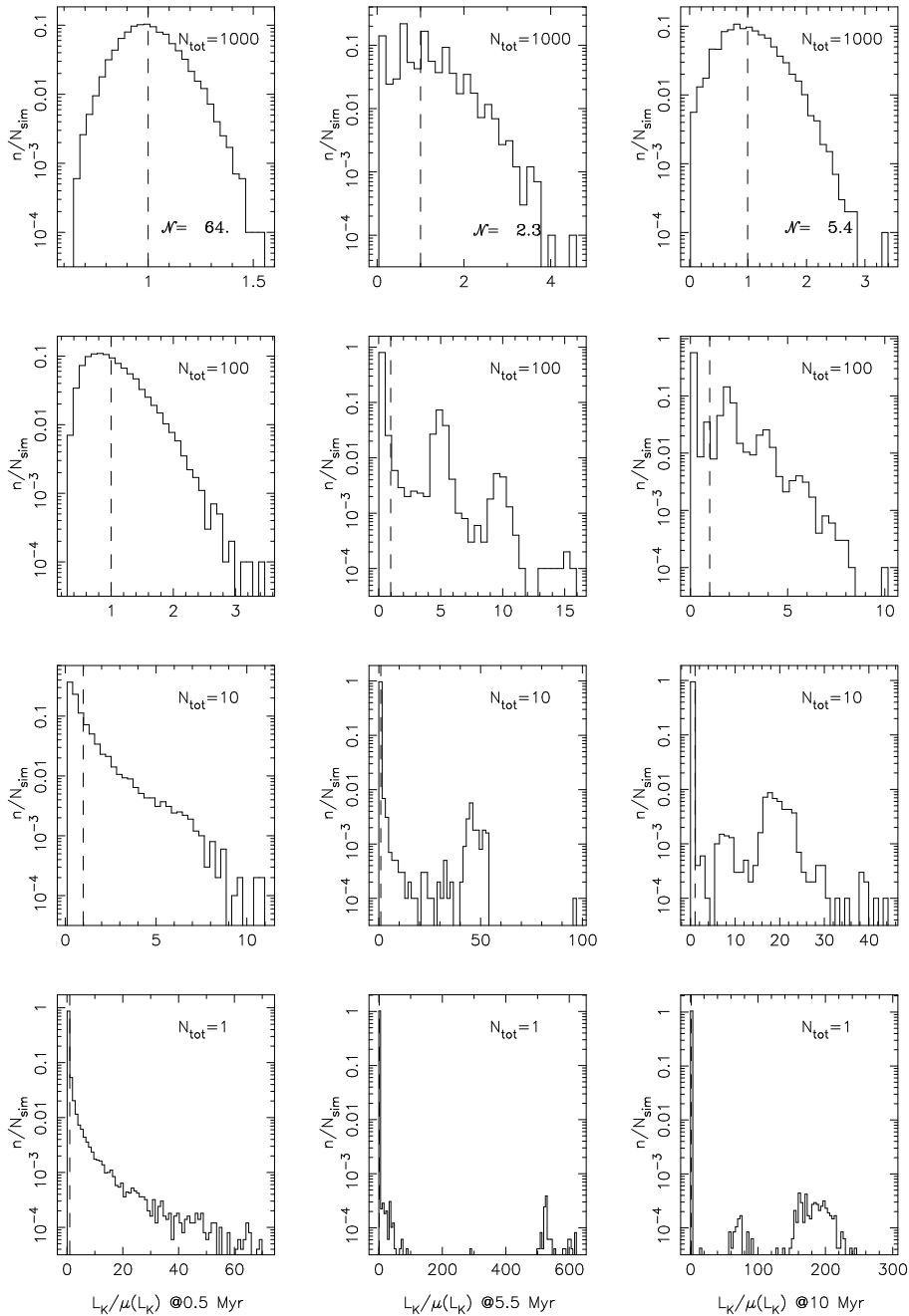} \caption{Same
 as Fig. \ref{fig:PDFlv} for $x=L_K/\mu(L_K)$.}
\label{fig:PDFlk} 
\end{figure*} 
 
In general, a linear combination of Poisson variables will not be another
Poisson variable, unless the weights take integer values \cite[see][ for a
discussion]{Cer02}. However, the fact that it is a {\it linear} combination
of random variables allows one to estimate {\it exactly} its first two
moments, even though the remaining moments can sometimes be also determined
analytically. This is well illustrated by Fig. \ref{fig:lum} which shows
the comparison of the Monte Carlo simulations with the fully-sampled
results for the $L_V$ and $L_K$ observables.  The agreement is very good in
all cases, with small deviations when the number of stars in the cluster
decreases. Except in rare occasions, the analytical expressions obtained
via Eq. (\ref{eq:neff}) are within 3 per cent and 5 per cent of the Monte
Carlo ones. Not only the average values agree at this level (and hence are
{\it not} biased) but also the estimation of the dispersion is very good,
over a dynamical range of three orders of magnitude in the total number of
stars.  {\it Linear functions of the luminosity (such as luminosities
integrated in a photometric band, or ionising fluxes) are not biased.}
 
Remarkably, our Poissonian formalism is also able to predict the mean value
and dispersion for simulations of clusters which contain only one star.  In
this case, the IMF gives us the relative weight of the probability $w_i$ to
obtain a star with a given luminosity $l_i$. The mean value, taking each
cluster as a one realisation, is

\begin{equation} 
\mu_L=\sum_{i=1}^{N_{sim}} \; l_i \, w_i, 
\end{equation} 

\noindent with  dispersion 

\begin{equation} 
\sigma^2(L)=\sum^{N_{sim}} (l_i - \mu_L)^2 w_i =
 \sum^{N_{sim}} l_i^2 w_i - \mu_L^2 .
\end{equation} 
 
Eq. (\ref{eq:neff}) implies that when ${\cal N}$ is very small $\sum l_i^2
w_i >> \mu_L^2$, and so Poisson dispersions and Monte Carlo ones become
very similar.

The highly variable correlation coefficient between $L_V$ and $L_K$ is also
predicted correctly by the Poisson formalism, as shown on
Fig. \ref{fig:lum}. The agreement is in fact at the 2 per cent level for
most ages.  It shows that a (variable) fraction of stars contribute in
luminosity to both bands, and therefore the actual dispersions in these
luminosities must take into account this important term if errors are to be
evaluated properly. As \cite{Cer02} showed, the effect of not including
this term is to systematically underestimate the dispersion.

So far so good for the first two moments, but what about the actual shape
of the PDF of these luminosities? Could the deviations from the Poisson
formalism be accounted for deviations in the PDF? Figures \ref{fig:PDFlv}
and \ref{fig:PDFlk} give these PDFs at selected ages in terms of the
variable $x = L / <L> = L / \mu_L$.  Note the logarithmic scale on the
vertical axis.  The vertical dashed lines give the values of
$x=\mu^{\mathrm{Analytical}}_L/\mu^{\mathrm{Monte Carlo}}_L$, which, as
shown above, are very close to unity and hence are not biased.  It is
interesting to note that whereas the mean value of the simulations are well
reproduced, the {\it median} value and the {\it most probable} (mode)
values are shifted to the left in all cases. This effect must be taken into
account when the result of standard synthesis models are compared with
individual clusters.

At early ages most stars are close to the ZAMS, and hence the
mass-luminosity relation makes that the PDF in luminosity is very similar
to the PDF distribution in total mass (compare with Fig. \ref{fig:IMF1}).
As stars evolve, post-MS evolutionary stages are reached, and obviously the
longer the stage, the better sampled it is. This gives rise to secondary
peaks in the PDF, where stars can accumulate, leading to multi-modality in
the luminosity distribution.  The number of such stars will be smaller than
1 per cent of the stellar population of the cluster. When the total number
of stars is low, there is a non-zero probability that the cluster will have
no post-MS. So, for a given fixed number of stars, it is possible to have
clusters (sharing the same total number of stars) with or without post-MS
stars, a clear case of possible bimodality when the observable depends on
the presence (or otherwise) of post-MS stars. This effect was first pointed
out in the pioneering work by \cite{Chi88}.  We stress again that the
average values are not biased, and that the deviations from the Poisson
distribution are relatively small in terms of the first two moments
only. Obviously higher moments are far more sensitive to shorter
evolutionary stages, and hence more prone to sampling effects.

We can nevertheless quantify these deviations in terms of $\cal{N}$. A
Poisson distribution defined by a parameter smaller than ${\cal{N}}=5$, has
a probability larger than 1 per cent to have a number of effective sources
equal to zero.  As mentioned before, the number of effective sources is not
an actual number of stars, but rather a measure of the effect of
(under-)sampling in these clusters.  For $\cal{N}$ values between 5 and
0.1, the probability of having a cluster with zero effective sources
increases from 1 per cent to 90 per cent.  For even smaller values of
$\cal{N}$ most clusters will have no effective sources in that band,
because they do not contain stars in the suitable evolutionary phase which
contributes to the observable.  We therefore expect that when $\cal{N}$ is
in the range between about 5 and 0.1 bimodality will be most easily
observed.

Again, $\cal{N}$ is an observable-dependent quantity, so it is possible
that some clusters have large $\cal{N}$ values for some quantities but
small values for others. If these quantities are linear functions of the
SED or the luminosity, then their average values are not biased. Any
non-linear function will, on the contrary, be prone to biases and may
enhance multimodality, as we show in the following sections.

\subsection{Logarithmic functions} 
 
\begin{figure*} 
\centering \includegraphics[angle=270,width=15cm]{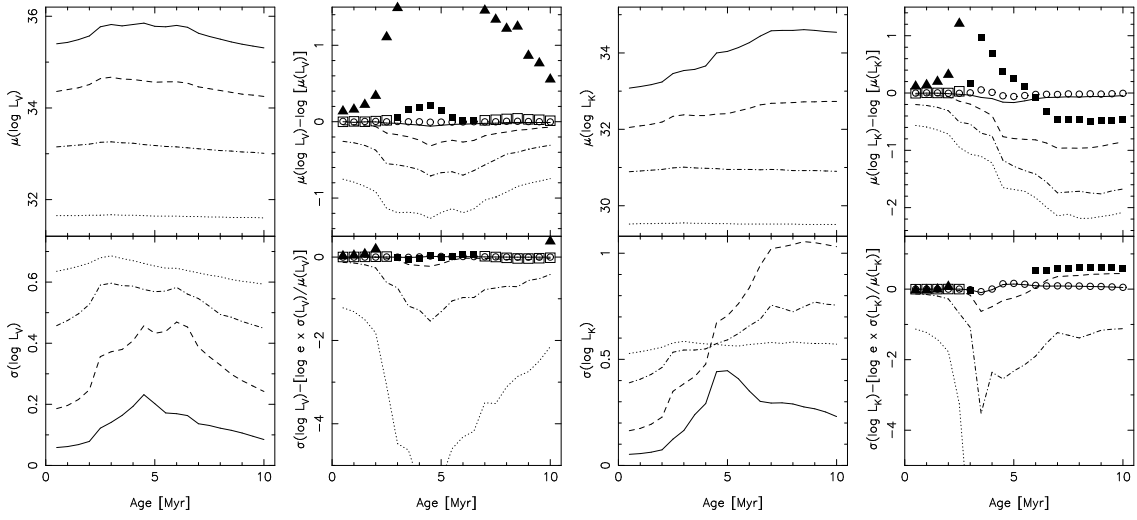}
 \caption{Comparison of average values and dispersions of $\log L_V$ and
 $\log L_K$ from Monte Carlo simulations and from Eqs. (\ref{eq:MeanLog})
 and (\ref{eq:siglog}) (used in synthesis models with an infinite number of
 stars).  The line styles and symbols correspond to different sizes of
 stellar populations ($\Ntot$=10$^3$ stars : solid lines and circles,
 $\Ntot$=10$^2$ stars : dashed lines and squares, $\Ntot$=10 stars :
 dot-dashed lines and triangles and $\Ntot$=1 star : dotted).  Symbols
 correspond to the use of Eq.  (\ref{eq:magcor}).  Filled symbols
 correspond to ${\cal N}(L)$ values lower than 1.}
\label{fig:loglum} 
\end{figure*} 

\begin{figure*} 
\centering \includegraphics[angle=270,width=12cm]{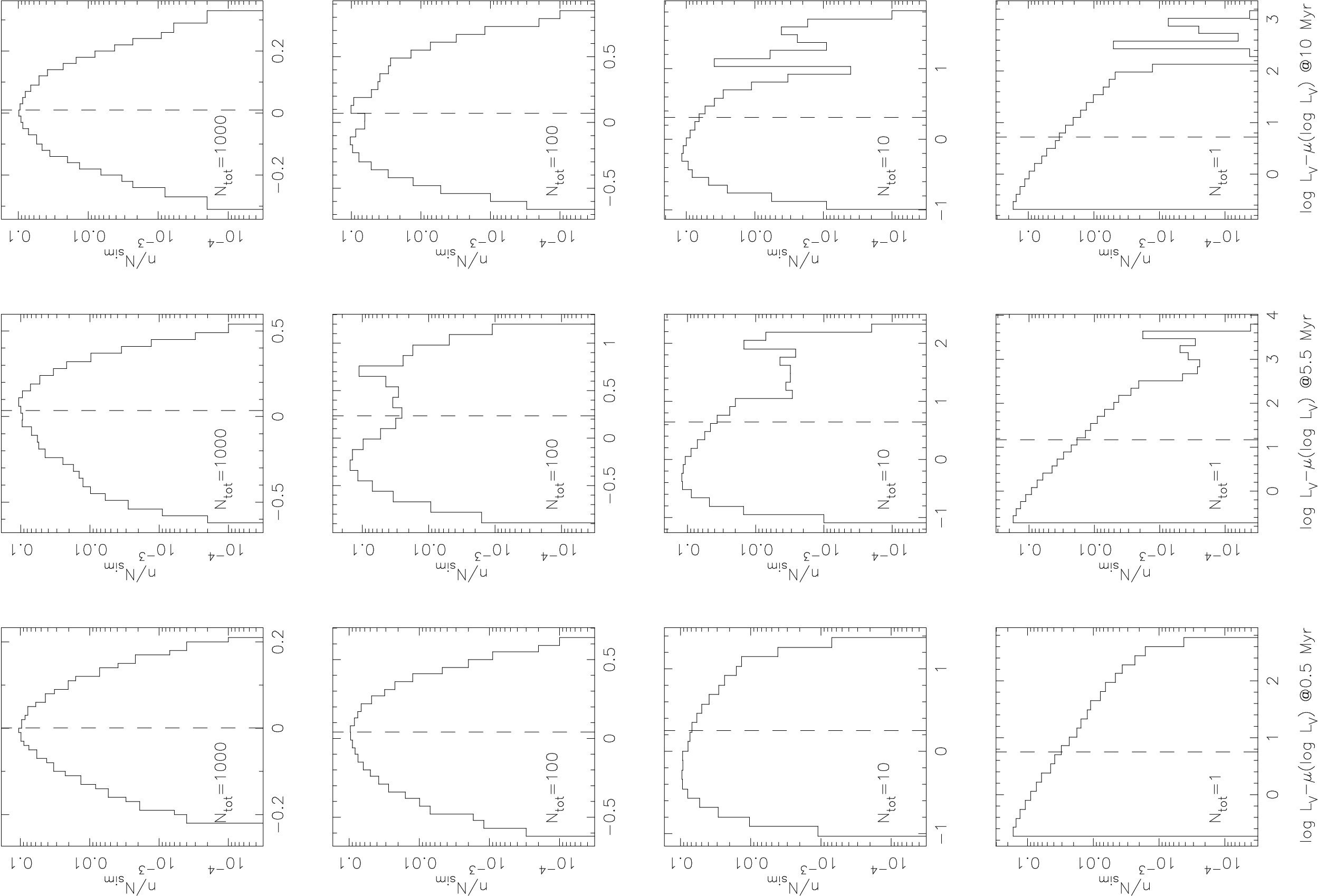}
\caption{PDF of $\log L_V$ at selected ages. The vertical dashed
line gives the position of $\log <L_V>$ and hence of the
bias. Note the variable amount of bias as a function both of age and
size of the stellar population.}  
\label{fig:PDFloglv} 
\end{figure*} 
 
\begin{figure*} 
\centering \includegraphics[angle=270,width=12cm]{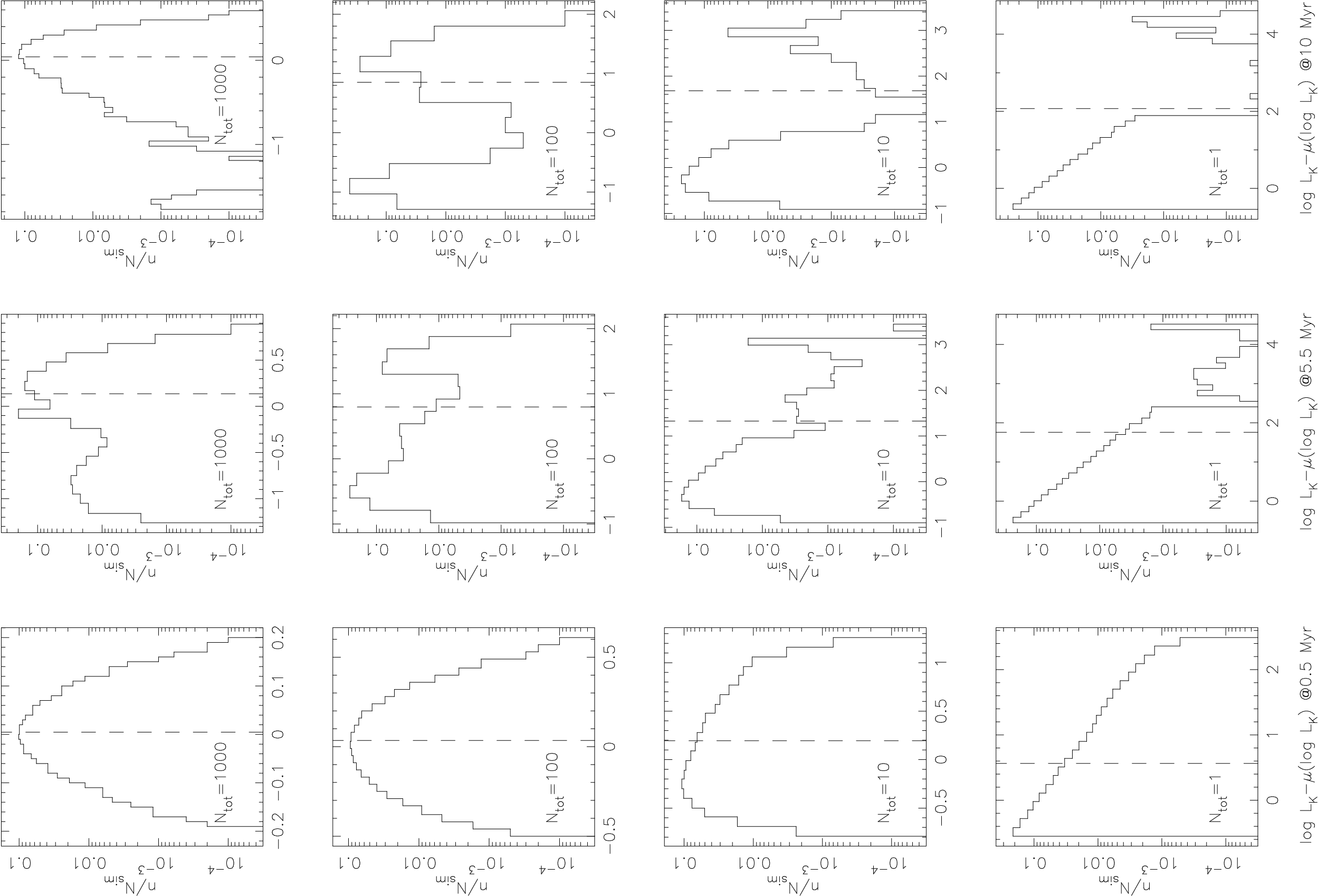}
\caption{PDF of $\log L_K$ at selected ages. The vertical dashed
line gives the position of $\log <L_K>$ and hence of the
bias. Note the variable amount of bias as a function both of age and
size of the stellar population. }  
\label{fig:PDFloglk} 
\end{figure*} 

In general, standard synthesis codes compute logarithmic quantities
assuming implicitely that their average value is the logarithm of the
average.  For the logarithm of the luminosity this would be for instance
 
\begin{equation} 
\mu_{\log L} = <\log L> 
\label{eq:MeanLog}
\end{equation}

\noindent and the corresponding dispersion would be evaluated as
 
\begin{equation} 
\sigma^2(\log L) = (\log {\mathrm e})^2 \frac{\sigma_L^2}{\mu_L^2} = 
\frac{(\log {\mathrm e})^2}{{\cal N}(L)} .
\label{eq:siglog} 
\end{equation} 

This is obviously not correct in general, since the assumption relies 
implicitely on an infinitely narrow-peaked PDF for $L$, which is seldom
the case. 

Let us assume a sample of observations of a definite-positive random
variable $L$ described by a PDF ${\cal P}(L)$. The expectation value
$\mu_C$ of any function of this variable $C=C(L)$ is
 
\begin{equation} 
\mu_C \; = \;  < C > \; = \; \int_{0}^\infty \, C(L) \, {\cal P}(L) \, dL .
\end{equation} 

Take for instance the case where $L$ is the luminosity in some photometric
band. The corresponding magnitude is $M=a+b\ln L$, with $a$ some zero point
and $b=-2.5 \log_{10} \mathrm{e}$. Its average value is

\begin{eqnarray} 
\mu_M = \; < M > &=&\int_{0}^\infty M \, {\cal P}(L)\, dL \nonumber \\ 
   & =&a\, +\, b\,\int_{0}^\infty (\ln L) \, {\cal P}(L)\, dL .
\label{eq:mulogV} 
\end{eqnarray} 
 
With the new variable $x = L / \mu_L$, whose PDF is ${\cal G}(x)$ such that 
of  course $\int {\cal G}(x) dx = \int {\cal P}(L) dL = 1$, this results in  

\begin{eqnarray} 
\mu_M&=&a\, +\, b\,\ln \mu_L \,  
   + \, b\, \int_{0}^\infty (\ln x) \,{\cal G}(x)\, dx  .
\label{eq:mulogV2} 
\end{eqnarray} 
 
Therefore, unless ${\cal G}(x) = \delta(x-1)$ (where $\delta(x)$ is Dirac's
Delta function) --that is, when the PDF of $L$ is defined {\it only} at
$\mu_L$ -- the average of the logarithmic quantity will not be the
logarithm of the average value of the variable.  The proper computation of
the ${\cal G}(x)$ functions can be only performed with a massive number of
Monte Carlo simulations. From this properly-sampled ${\cal G}(x)$ one could
extract samples of $x$ for any sample of size $\Ntot$, but this is
obviously unpractical. If this is nevertheless not done, then the estimated
values are very likely to be biased.

A rough estimate of the bias can be evaluated by expanding in Taylor
series.  For the magnitude $M$ this is

\begin{equation}
M=a+b \ln \mu_L + b\left(\frac{L-\mu_L}{\mu_L}
  -\frac{(L-\mu_L)^2}{2}\frac{1}{\mu^2_L} + \cdots \right),
\end{equation}

\noindent so that, to first order, the mean value of the magnitude, and its
variance are

\begin{eqnarray} 
\mu_M&=&a\, +\, b\,\ln \mu_L \, - \, b\, \frac{\sigma^2_L}{2 \mu^2_L} +
...     \nonumber \\
     &=&a\, +\, b\,\ln \mu_L \, - \, \frac{b}{2 {{\cal N}}(L)} +... \\
\sigma^2_M&=&\frac{b^2 \sigma^2_L}{\mu^2_L}
\left(1-\frac{1}{4}\frac{\sigma^2_L}{\mu^2_L}+...\right)\nonumber \\
 &=&\frac{b^2}{{\cal N}(L)}-\left(\frac{b}{2 {\cal N}(L)}\right)^2 +...
\label{eq:magcor}
\end{eqnarray} 

To first order, the bias is then $\frac{b}{2 {\cal{N}}(L)}$, so if
${\cal{N}}(L)$ is large enough, Eqs. (\ref{eq:MeanLog}) and
(\ref{eq:siglog}) are good approximations. In many cases ${\cal{N}}(L)$
will be small and the bias important. Eq. (\ref{eq:magcor}) gives a lower
limit to the bias (since higher order corrections were not taken into
account) and the only way to estimate the bias is, inevitably, via Monte
Carlo simulations. This is done in Fig. \ref{fig:loglum} where this
analytical estimate is compared with the Monte Carlo ones (middle and right
panels).

Alternatively, in terms of difference in magnitude

\begin{equation}
{\Delta M_L}=\mu(M_L)-M_L= 2.5 (\log L - \mu_{\log L}),
\end{equation}

\noindent we see that the bias can reach some 0.5 magnitudes in $M_V$ and
2.5 magnitudes in $M_K$ for clusters with 100 stars between 2 and 120
$M_\odot$, and 2 magnitudes in $M_V$ and $\sim 5$ magnitudes in $M_K$ for
groups of 10 stars in that mass range. We can also see that using Eqs.
(\ref{eq:MeanLog}) and \ref{eq:siglog} give systematically wrong results,
because they predict systematically too large values. This is easy to
understand since they assumed a fully-sampled IMF, were massive stars are
(comparatively) over represented with respect to Monte Carlo samples of
small size. The estimates given by Eq. (\ref{eq:magcor}) are indicated with
symbols in Fig. \ref{fig:loglum}. Although they fail to give the correct
correction (due to the lack of higher order terms) in the bias, they
provide a rough indication of the underestimation in the dispersion.

Figs. \ref{fig:PDFloglv} and \ref{fig:PDFloglk} present the PDFs of $\log
L_V$ and $\log L_K$ respectively.  In clusters with ${\cal{N}}(L_V)$ or
${\cal N}(L_K)$ larger than 10 (see Figs. \ref{fig:PDFlv} and
\ref{fig:PDFlk}), the PDFs are unimodal, as expected.  However, for smaller
values of ${\cal{N}}$ the situation is entirely different.  In this case,
the cluster is dominated by main sequence stars, which do not sample
properly the IMF, given their small numbers, and in particular the upper
mass end.

At 5.5 and 10 Myr, the $L_V$ distribution of clusters with 100 stars have
${\cal{N}}(L_V)$ values of 0.7 and 2.5 respectively (see
Fig. \ref{fig:PDFlv}). The associated PDFs of $\log L_V$ show a bimodal
structure with two peaks of similar amplitude (Fig. \ref{fig:PDFloglv}).
The effect is more extreme in the case of $\log L_K$
(Fig. \ref{fig:PDFloglk}) were bimodality is present even for clusters with
1000 stars (in the 2--120 $M_\odot$ mass range) at 5.5 Myr with
${\cal{N}}(L_K)=2.3$ (see also the PDFs at 5.5 and 10 Myr for clusters with
100 stars, with ${\cal{N}}(L_K)$ values of 0.23 and 0.54
respectively). Bimodality with peaks of roughly similar amplitude implies
that real clusters should have (log)luminosities peaking around (roughly)
two values (barring selection effects, etc). One could naively think that
these are two distinct populations of clusters where in fact there is only
one. Since the total number of stars is the same, one could be driven to
conclude, mistakenly, that the IMF in these two 'populations' is different,
while in fact it is just a sampling effect.

A final comment on simulations of clusters with one star, because their
PDFs help to understand the origin of multimodality and bias.  At 0.5 Myr
the PDF is roughly a power law (the convolution of a IMF power law with the
power-law relation between mass and luminosity for main sequence
stars). Figs.  \ref{fig:PDFloglv} and \ref{fig:PDFloglk} indeed show a
linear relation in these log-log plots.  As time goes by, the more luminous
stars (i.e. the most massive ones) leave the main sequence, and become more
luminous than the brighter main sequence stars. The PDFs show two distinct
r\'egimes: (i) a power-law for the less massive (and thus dimmer) stars,
and (ii) secondary peaks for the longer post-MS evolutionary phases. When
more massive clusters are considered, these post-MS peaks grow in
importance, become better sampled, and eventually the bias disappears.

\subsection{Rational and log-rational functions} 

\begin{figure*}
\centering \includegraphics[angle=270,width=12cm]{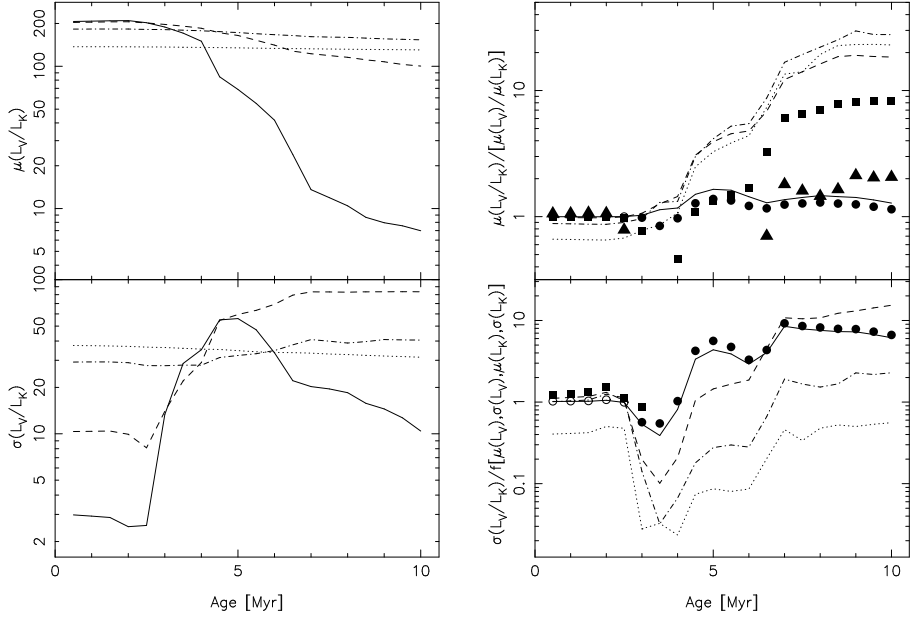}
 \caption{Mean values and dispersions of $L_V/L_K$ in
 Monte Carlo simulations and in fully-sampled synthesis (via
 Eqs. (\ref{eq:meanratio}) and (\ref{eq:sigratio}), as used in standard
 synthesis models).  Symbols and lines as in Fig. \ref{fig:loglum} except
 for filled symbols, that now correspond to ${\cal N}(L_V)$ and
 ${\cal N}(L_K)$ values lower than 10. Note that the $\sigma$
 value for clusters with $10^3$ stars is larger at 5 Myr than the $\sigma$
 value for 'clusters' with one star. It is due to the PDF
 (Fig. \ref{fig:PDFratio}) of the clusters with one star, see text for
 explanations.}
\label{fig:ratio} 
\end{figure*} 

\begin{figure*} 
\centering \includegraphics[angle=270,width=12cm]{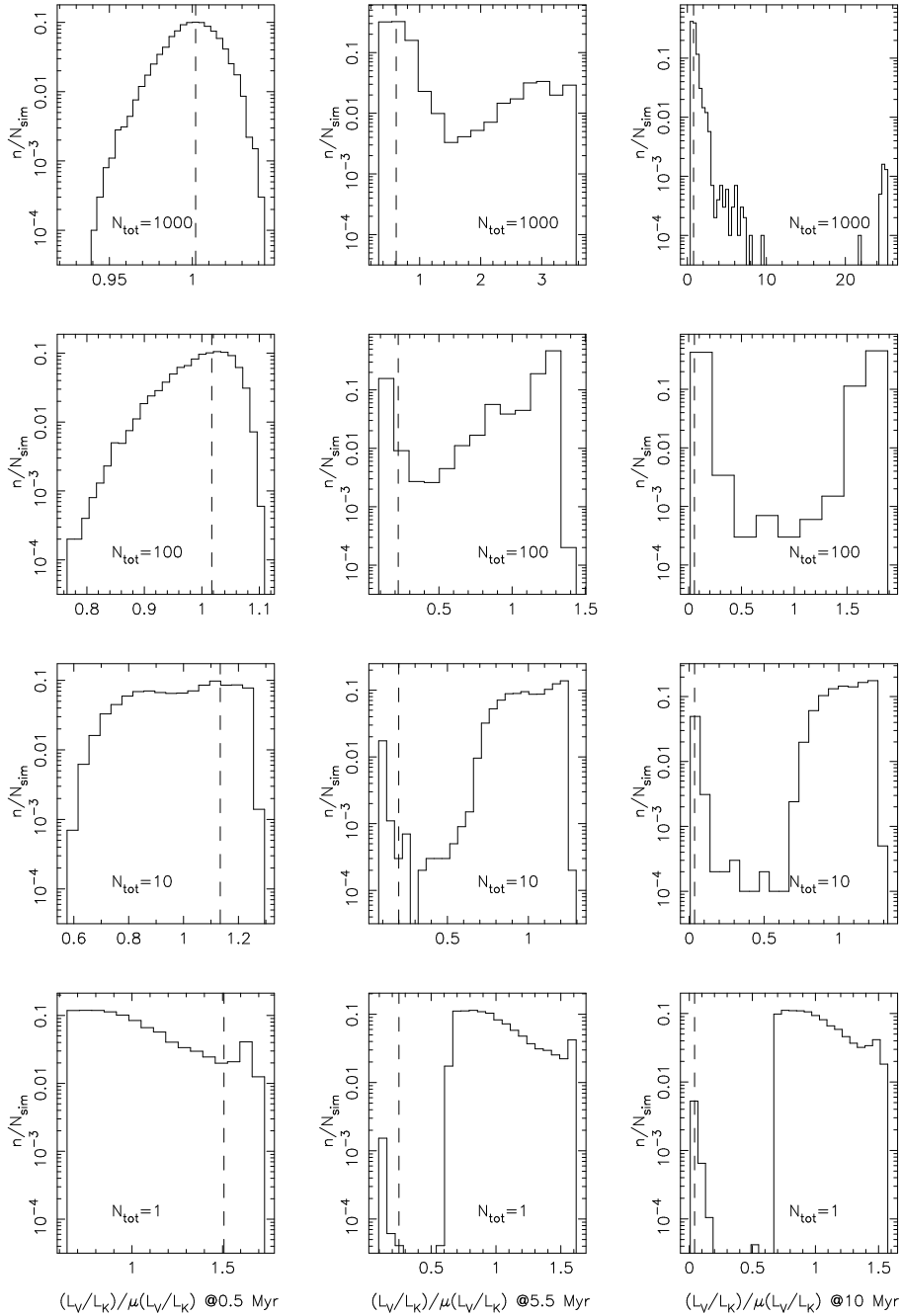}
\caption{PDF at selected ages of the $L_V/L_K$ ratio. The
standard (naive) theoretical values are shown with vertical dashed lines. }
\label{fig:PDFratio} 
\end{figure*}   
 
Rational functions of luminosities are even more important in the
interpretation of the properties of small stellar systems because, naively,
they seem to be independent of normalisation problems, since the usual
scaling with mass or number of stars cancels out.  But this may not be
necessarily the case, and, in addition, their dispersion due to the uneven
sampling of the IMF can be dramatically large.

In standard synthesis models it is assumed that for a rational function 
$u=x/y$ its average value is given by

\begin{equation} 
\mu_u \; = \; \mu_{x} \times \mu_{1/y} \; = \; \mu_x / \mu_y ,
\label{eq:meanratio} 
\end{equation} 

\noindent and its dispersion is given by:

\begin{eqnarray} 
\frac{\sigma_u^2}{\mu^2_u} &\simeq& \frac{\sigma_x^2}{\mu^2_x} +
\frac{\sigma_y^2}{\mu^2_y} -  
2 \frac{{\mathrm{cov}}(x,y)}{\mu_x\, \mu_y} \nonumber \\ 
&=& \frac{1}{{\cal N}(x)} + \frac{1}{{\cal N}(y)} - 2 
\frac{\rho(x,y)}{\sqrt{{\cal N}(x)\,{\cal N}(y)}} = \frac{1}{{\cal N}(u)} ,
\label{eq:sigratio} 
\end{eqnarray} 

\noindent where $\rho(x,y)=\frac{\mathrm{cov}(x,y)}{\sigma_x\,\sigma_y}$ is
the correlation coefficient between $x$ and $y$.

Again, this would be correct if the distribution of $u$ would be a very
narrow peak, but in the general case this is not correct.  For a function
$R=1/L$, the average value is given by
 
\begin{equation} 
\mu_R \; = \; \int_{0}^\infty  
 \, \frac{1}{L} {\cal P}(L)\, dL = \; \frac{1}{\mu_L} \int_{0}^\infty   
 \frac{1}{x} \, {\cal G}(x) \, dx ,
\end{equation} 

\noindent where $x = L \, / \mu_L$.  Again, 
unless  ${\cal G}(x) = \delta(x-1)$, 
 
\begin{equation} 
\mu_R \; = \; \mu_{1/L} \; \neq \; 1 / \mu_L ,
\end{equation} 

\noindent and, more generally,  
\begin{equation} 
 \mu(L_1/L_2) \; \neq \; \mu_{L_1} / \mu_{L_2} .
\end{equation}

With a second-order Taylor expansion one gets, to first order

\begin{eqnarray} 
\mu_{R} & = & \frac{1}{\mu_L} \left(1+\frac{\sigma^2_L}{\mu^2_L}+...\right)
 = \frac{1}{\mu_L} \left(1+\frac{1}{{\cal{N}}(L)}+...\right)\nonumber \\
 \sigma^2_R & = &\frac{\sigma^2_L}{\mu^4_L}
 \left(1-\frac{\sigma^2_L}{\mu^2_L}+...\right) \nonumber \\ & = &
 \frac{1}{\mu^2_L \,{\cal{N}}(L)} \left(1-\frac{1}{{\cal{N}}(L)}+...\right),
 \nonumber \\
\end{eqnarray} 

\noindent and, in general, for a rational function $u=x/y$

\begin{eqnarray} 
\mu_{u} & = & \frac{\mu_{x}}{\mu_{y}}
\left(1+\frac{\sigma^2_{y}}{\mu^2_{y}}\right) -
\frac{\mathrm{cov}(x,y)}{\mu^2_{y}}+... \nonumber \\
  &=&\frac{\mu_{x}}{\mu_{y}}
\left(1+\frac{1}{{\cal{N}}(y)}-
\frac{\rho(x,y)}{\sqrt{{\cal N}(x)\,{\cal N}(y)}}\right) +... \nonumber \\
\sigma^2_{u} & = &\frac{\sigma^2_{x}}{\mu^2_{y}} +
\frac{\mu^2_{x}\,\sigma^2_{y}}{\mu^4_{y}} - \frac{2\,\mu_{x}
  \mathrm{cov}(x,y)}{\mu^3_{y}} - \nonumber \\
&&\left(\frac{\mu_{x}\,\sigma^2_{y}}{\mu^3_{y}}
- \frac{\mathrm{cov}(x,y)}{\mu^2_{y}}\right)^2 + ...\nonumber \\
\frac{\sigma_u^2}{u^2} &=& \frac{1}{{\cal N}(x)} + \frac{1}{{\cal N}(y)} - 2 
\frac{\rho(x,y)}{\sqrt{{\cal N}(x)\,{\cal N}(y)}} -\nonumber \\ 
& &  \left(\frac{1}{{\cal N}(y)} - 
\frac{\rho(x,y)}{\sqrt{{\cal N}(x)\,{\cal N}(y)}}\right)^2 +...
\label{eq:Rcor}
\end{eqnarray}

To first order approximation, the bias is then of the order of
$\frac{\mu_{x}}{\mu_{y}}\left(\frac{1}{{\cal N}(y)} -
\frac{\rho(x,y)}{\sqrt{{\cal N}(x)\,{\cal N}(y)}}\right)$, and the correct
estimation of the mean value of a ratio requires the covariance between the
numerator and denominator.

Again, we illustrate here these effects with the simple ratio $L_V/L_K$ in
Fig.  \ref{fig:ratio}, where Eqs.  (\ref{eq:meanratio}) and
(\ref{eq:sigratio}) produce a very severe bias, even larger than one order
of magnitude, both in the average and in the dispersion.  The symbols show
the results using Eq. (\ref{eq:Rcor}) and they have the same meaning as in
Fig. \ref{fig:loglum} except for filled symbols, which now correspond to
${\cal N}(L_V)$ and ${\cal N}(L_K)$ values lower than 10. In this case, the
use of the Eq. (\ref{eq:Rcor}) only produces a marginally better result
than Eq. (\ref{eq:meanratio}). Note that in the case of $\sigma(L_V/L_K)$
there are fewer points computed with Eq. (\ref{eq:Rcor}) than in the case
of the mean value. This is just because, for some values,
Eq. (\ref{eq:Rcor}) produces negative $\sigma^2(L_V/L_K)$ values and the
use of these equations is not valid.  It is interesting to see that the
bias can be positive or negative, a direct consequence of the time
variability of the correlation coefficient between the two luminosities.

The corresponding PDFs are given in Fig. \ref{fig:PDFratio}.  In this case,
bimodality is apparent for a wide range of situations, and again, it is due
to a secondary peak with larger values of the $L_V/L_K$ ratio.  The bias
may increase or decrease the average value, as shown by the vertical dashed
lines. It is remarkable to see that the actual dispersion for the more
massive clusters, at 10 Myr for example, is actually larger than the
dispersion in less massive clusters.  Again this is due to the presence of
very massive stars, better sampled in these massive clusters, which have in
comparison a much smaller flux in the $K$ band, and hence a very large
$L_V/L_K$ ratio.  The behaviour of the bias with the number of stars is
clearly non-monotonic, as shown also in Fig.  \ref{fig:ratio}.\\

Finally, in the case of log-rational functions (such as integrated colours)
there will be a priori a double bias, arising from the use of two
non-linear functions.  And yet most codes will still assume that the
colours of a stellar population will not depend on its total mass.  We take
again as an example the $V-K$ colour. To first order,

\begin{eqnarray}
\mu_{V-K}&=&a\, +\, b\,\ln \frac{\mu_{L_V}}{\mu_{L_K}}
\nonumber \\ 
& & - \, \frac{b}{2}\left(\frac{1}{{\cal N}(L_V)} - \frac{1}{{\cal
      N}(L_K)} \right) +...\nonumber \\
\sigma^2_{V-K}&=&b^2\left(\frac{1}{{\cal N}(L_V)} + \frac{1}{{\cal
      N}(L_K)} - 2\frac{\rho(x,y)}{\sqrt{{\cal N}(L_V)\,{\cal N}(L_K)}}\right) - \nonumber \\ 
& &  \, \frac{b^2}{4}\left(\frac{1}{{\cal N}(L_V)} - \frac{1}{{\cal
      N}(L_K)} \right)^2 +...
\end{eqnarray}

Note that in this case the bias does not dependent on the correlation
coefficient, and it has a value of approximately
$\frac{b}{2}\left(\frac{1}{{\cal N}(L_V)} - \frac{1}{{\cal N}(L_K)}
\right)$.  Since redder wavelengths have a lower $\cal{N}$ value (in this
range of ages) than shorter wavelengths \cite[see][ for details]{Cer02},
this approximation for the bias indicates that predictions of synthesis
models must be corrected with a blue shift in order to get back to the
actual {\it mean} value of a set of observed clusters. Note that for older
clusters (say, above 1 Gyr or so), the trend is reversed since redder
wavelengths (cooler evolutionary phases) become better sampled.

\begin{figure*} 
\centering \includegraphics[angle=270,width=15cm]{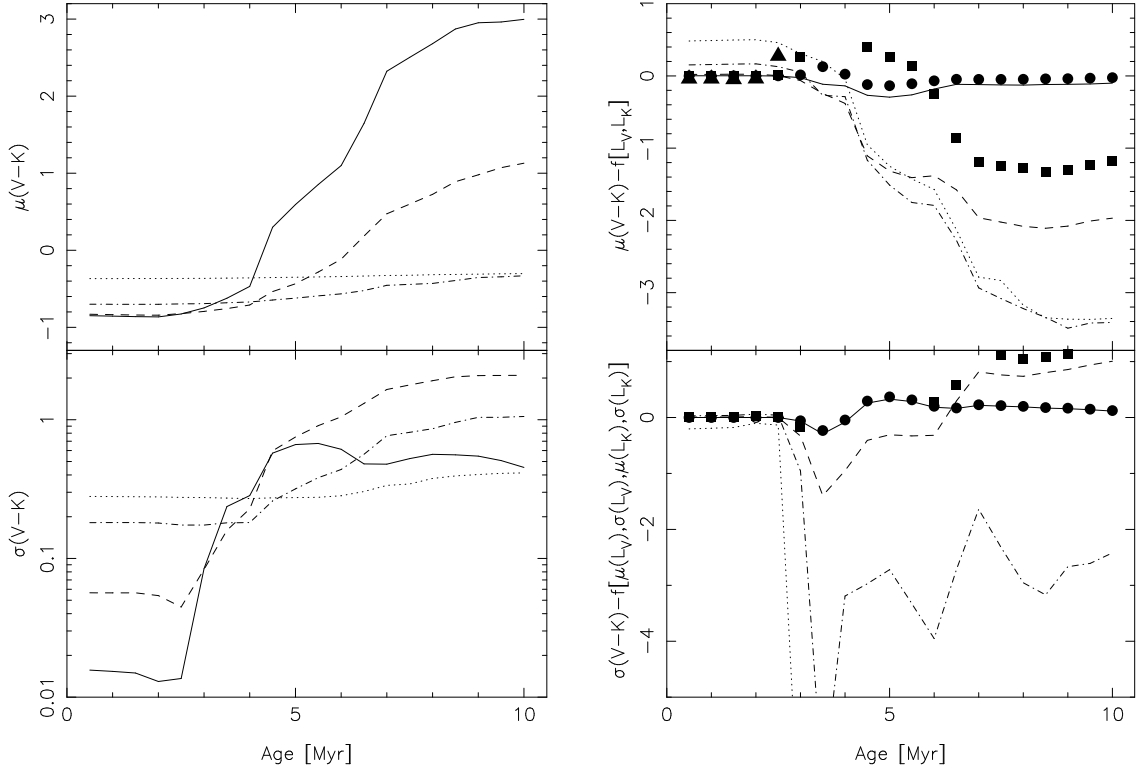}
 \caption{Mean values and dispersions of the integrated $V-K$ colour, from
 Monte Carlo simulations, and their comparison to simulation results
 obtained directly from the corresponding band
 luminosities. Symbols as in Fig. \ref{fig:ratio}.}
\label{fig:VK} 
\end{figure*}

\begin{figure*} 
\centering \includegraphics[angle=270,width=12cm]{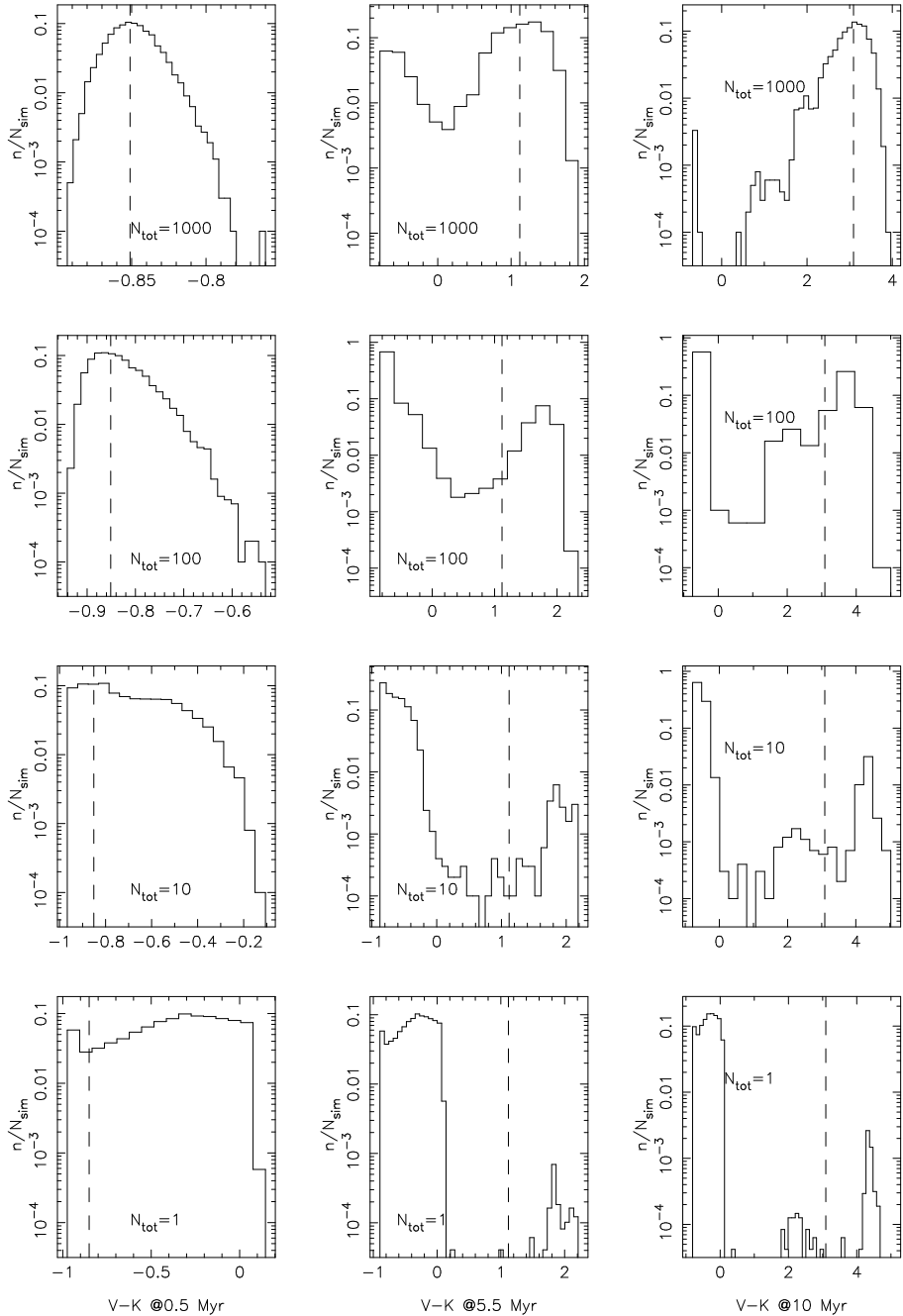} 
 \caption{PDFs at selected ages for the $V-K$ colour. 
Naive theoretical values are shown with short-dashed vertical lines.} 
\label{fig:PDFVK} 
\end{figure*}

The situation is illustrated in Figs. \ref{fig:VK} and \ref{fig:PDFVK}. The
top right panel of Fig. \ref{fig:VK} shows the mean $V-K$ colour for the
different sets of simulations. In this case, the difference between actual
colours and those predicted by fully-sampled synthesis models can reach
more than 3 magnitudes, even for the most massive clusters.

In the case of young star forming regions (younger than some 20 Myr), red
colours have lower $\cal{N}$ values than blue ones \cite[see][ for
details]{Cer02}. The resulting bias shows that the predictions of synthesis
models must be corrected toward blue colours in order to reproduce the {\it
mean} value of a set of observed clusters. At older ages, the bias may turn
to bluer or redder values depending on the age, that is, on the proportion
of post-MS versus main sequence stars.

Another interesting result shown in Figs. \ref{fig:ratio} and \ref{fig:VK}
is the corresponding $\sigma$ value. In the case of Gaussian distributions,
$\sigma$ corresponds to the half width of maximum at about 60 per cent of
the full height. In our case, the PDFs are not Gaussians, but $\sigma$
still gives us the width of the distribution at some (undefined)
height. The Monte Carlo simulations show that at some ages (when post-MS
stars appear in the clusters), the $\sigma$ value is maximum for the
simulations with 100 stars and decreases for even fewer stars. This value
is also related with the area covered by the simulations in
Fig. \ref{fig:GTC}, where the 'dispersion' in the diagram is maximum for
simulations with 100 stars. As we have already seen, this maximum $\sigma$
value is also related with a larger probability of obtaining bimodal
distributions. The reason of this behaviour is that most of the clusters
with a smaller number of stars will not have the post-MS phases sampled at
all. This is also consistent with the lack of evolution in the $L_V/L_K$
ratio and the $V-K$ colour. So, the probability of finding a cluster with
post-MS falls below the width of the level defined by $\sigma$. For more
populated clusters, almost all of them will be reproduced correctly by
standard synthesis models with a non-zero number of post-MS stars.
  
It is worth emphasizing that this behaviour is also present in simulations
performed by \cite{Bru02Tuc}. In his diagrams there is a maximum dispersion
for the simulations with 10$^4$ $M_\odot$ that decreases for simulations
with 10$^3$ $M_\odot$. Also, the bias effect is clearly present in his
Monte Carlo simulations since the colours of these simulations do not agree
with his analytical predictions.

\section{Discussion and implications}

\begin{figure*} 
\centering \includegraphics[angle=270,width=12cm]{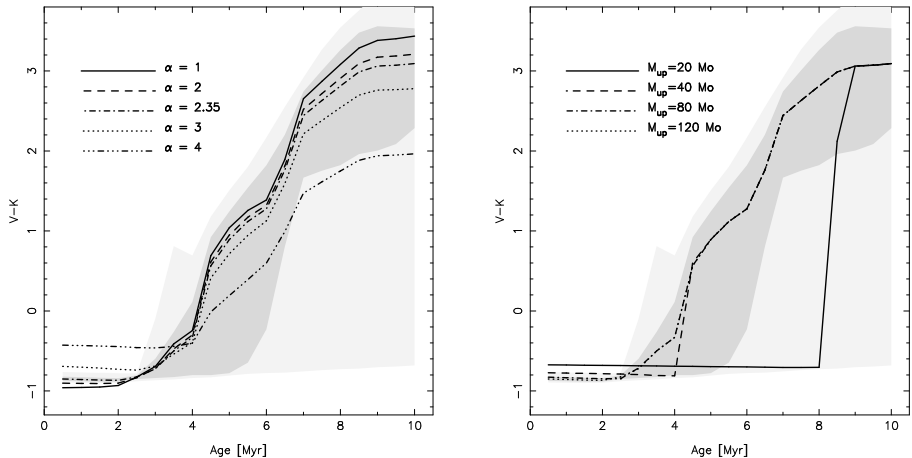} 
\caption{Evolution of the $V-K$ colour for different IMF slopes (left,
assuming that $M_\mathrm{up}$= 120 M$_\odot$) and M$_\mathrm{up}$ values
(right, assuming a Salpeter IMF slope). The areas in dark gray shade
corresponds to the 90 per cent confidence interval for clusters with 1000
stars. The areas in light gray shade corresponds to the maximum and minimum
values obtained in simulations of clusters with 1000 stars.}
\label{fig:VKIMF} 
\end{figure*} 

The comparison of observations with the results of a population synthesis
method is in fact an inverse problem : Given the data, what are the models
which best fit the data, given the errors in the data? It is seldom the
case that errors within the models are also taken into account. It is also
important to analyse whether the best fitting models are also good fits,
and whether the solution is unique or not. In the case of synthesis models
this procedure is often neglected and a variation in the input parameters
is much preferred.

We have shown in the previous sections two major problems. First, the usual
scaling with the total mass of the stellar population should not be made
when dealing with small systems (dwarf galaxies, starbursts, stellar
clusters, pixels, etc). Second, observables which are non-linear functions
of the SED (or equivalently of the monochromatic luminosity) may present
not only multimodal distribution functions, but also a systematic bias in
the average values of these observables, with respect to the values
predicted by codes which fully sample the IMF.

These effects have been overlooked in the past, because most synthesis
codes do not provide an estimate of the intrinsic dispersion of the
observables. Whilst this was easy to understand when dealing with the
integrated properties of massive stellar populations, such as spiral and
elliptical galaxies, this becomes a serious problem if such codes are
applied to smaller stellar systems. Formally, the first moment of the PDF
of a given observable cannot be scaled down properly. A possible solution
would be to have the PDF of the observable and then extract samples of
small size from that PDF. In this way the statistical properties of the
observable, in small samples, can be analysed and compared with the
observations.

The only proper solution to deal with the two problems we have discussed in
this paper, multimodality and bias, is therefore to perform Monte Carlo
simulations to construct the PDF of the observable. This is often
unpractical, and we have shown that the quasi-Poisson formalism that we
have developed can account for the statistical properties of the first two
moments at a very good approximation level. Although the formalism cannot
account for higher order moments, and thus to allow for strong deviations
in the PDFs, it does not suffer from systematics and can easily (and
should) be included in any synthesis code.

As we mentioned before, it has often been the case that the easiest
solution to account for 'peculiar' observations was to change some of the
input parameters of the synthesis code, and in particular the properties of
the IMF. It is well beyond the scope of this paper to analyse in detail
each and every claim for variations of the IMF in small stellar systems,
but we will illustrate here how sampling effects may have been mistakenly
overlooked. We take again the $V-K$ colour, a non-linear function of the
SED, as an example. Figure \ref{fig:VKIMF} (right panel) shows its
evolution with time when the upper mass limit of the IMF is changed. It is
readily apparent that if an observation gives a very blue $V-K$ colour, and
all other parameters are the same (such as mass, luminosity, etc) within a
small range, then it is tempting to hastily conclude that the upper mass
limit in that particular system has to decrease.  But in fact, as
Fig. \ref{fig:VK} shows, not only the average colour changes with the mass
of the system, but also a large dispersion is expected, arising {\it only}
from sampling fluctuations.  Similarly, Fig. \ref{fig:VKIMF} (left panel)
shows the effect of varying the slope of the IMF. Again, a too blue colour
may be interpreted as evidence for a steeper slope.  While our results
cannot confirm or reject possible IMF variations in these systems, they
however clearly indicate that part (if not all) of the dispersion in the
$V-K$ colour may be accounted for by the incomplete sampling of a single,
universal IMF.

Two final comments. First, in some cases there is no need for a bias to be
present to explain some peculiar properties of small stellar systems. All
the values quoted in Fig. \ref{fig:VKIMF} lie in fact in the predicted band
quoted by the maximum and minimum $V-K$ values obtained from the
simulations of clusters with 1000 stars (where the bias is small). The very
large dispersion, an expected intrinsic property due to incomplete
sampling, may account for many of the observed values, with no need to
invoke special properties.  As an example, in high-metallicity regions with
Wolf-Rayet stars, the dispersion of synthesis models around their mean
value is enough to explain the observations, as shown by \cite{BK02}.

Second, another interesting effect concerns the ages of these systems, as
inferred from non-linear functions of the luminosity. Because these
observables may have a wide, multimodal distribution function, the
intrinsic dispersion which may be observed in a given sample is not
necessarily a reflection of a variation in the properties of individual
members of that sample. It appears very dangerous to infer ages or mass
variations from these small systems.

\section{Conclusions}

In this work we have found important effects which have been overlooked in
the application of population synthesis codes to small stellar systems such
as clusters, starbursts, dwarf galaxies or pixels.  Most current synthesis
codes predict the average values of observables for an infinitely large
population of stars, which samples perfectly a given IMF. Extrapolating
these predictions to small stellar systems may be misleading.

First, scaling observables with the mass in stars is not correct for small
systems.  The proper scaling should be performed with the total number of
stars present in the system.

Second, the distribution functions of observables which are non-linear
functions of the monochromatic luminosity (or SED) may show
multimodality. This implies that for a given population, of fixed size, and
all other parameters kept the same, there will be a large dispersion in the
observable, whose average value will not necessarily be the same as the one
predicted when the IMF is fully sampled. Most synthesis codes will
therefore lead to biased predictions when applied to small stellar systems.

Third, we have also shown that bimodality is a natural effect that appears
when stellar systems are affected by sampling (i.e. the number of stars in
the clusters is not enough to sample completely the IMF).  It is related to
the evolutionary speed of different masses, or in more graphical terms, to
the absence/presence of stars in some evolutionary phases.  This effect was
already shown 14 years ago by \cite{Chi88}.  We have now established the
range of $\cal{N}$ values (from $\approx$ 5 to 1) where bimodality appears.
An additional result is that bimodality will be more easily observed when
infrared colours are used since such colours are more affected by the
presence of short-lived post-MS phases.

Fourth, while we cannot confirm or reject the hypothesis of possible
variations in the properties of the IMF in small stellar systems, these
results suggest that an important fraction of the dispersion observed in
non-linear observables is not necessarily evidence for IMF variations, but
can rather be accounted for by sampling fluctuations.

Finally, we have also derived an operational limit to the validity of
standard synthesis codes. There is a critical effective number of stars
(unrelated to an actual number of stars), with a value of around 10, which
defines a boundary where multimodality may appear, i.e.  cluster masses
larger than 10$^5$ M$_\odot$ for a Salpeter IMF in the range 0.08 -- 120
M$_\odot$ (but the exact value for the cluster masses depends on the age
and the observable).  Below this critical value, the results of the codes
must be taken with caution, since not only they can be biased, but also
they may underestimate the actual dispersion of the observable, and Monte
Carlo simulations are required. The reason for the existence of this
critical value is directly related to the presence (or otherwise) of
stellar evolutionary stages which are not fully sampled in small systems.

We encourage the inclusion of the calculation of this effective number of
sources for a given observable in all synthesis codes, so that observers
(and theoreticians alike) can estimate the expected dispersion (and
possible biases) in observables.

\section*{Acknowledgments}

We thank Roberto Terlevich for useful comments on the
difference between distributions  plotted in linear and
logarithmic scales. We also thank Francisco Javier Castander,
Valentina Luridiana, Enrique P\'erez, Jos\'e V\'\i lchez and Gustavo
Bruzual for discussions on statistics and observational data. This
project has been partially supported by the AYA 3939-C03-01 program.

\label{lastpage}
 
\end{document}